\documentclass[fleqn,usenatbib]{mnras}

\usepackage{newtxtext,newtxmath}

\usepackage[T1]{fontenc}
\usepackage{ae,aecompl}

\usepackage{graphicx}	

\usepackage{soul} 
\usepackage{xargs} 
\usepackage[pdftex,dvipsnames]{xcolor}  
\usepackage{amsmath}
\usepackage{rotating}
\usepackage{pdflscape}
\definecolor{shade}{gray}{0.9}

\newcommand{\hst }{{\em HST}}
\newcommand{\gaia}{{\em Gaia}}
\newcommand{\cxo }{{\em CXO}}

\title[New magnetar counterparts from HST]{New candidates for magnetar counterparts from a deep search with the Hubble Space Telescope}

\author[A. A. Chrimes et al.]{A. A. Chrimes,$^{1}$\thanks{E-mail: a.chrimes@astro.ru.nl}
A. J. Levan$^{1,2}$,
A. S. Fruchter$^{3}$,
P. J. Groot$^{1,4,5}$,
C. Kouveliotou$^{6,7}$,
J. D. Lyman$^{2}$,\newauthor
N. R. Tanvir$^{8}$
and K. Wiersema$^{9}$\\
$^{1}$Department of Astrophysics/IMAPP, Radboud University, P.O. Box 9010, 6500 GL, The Netherlands \\
$^{2}$Department of Physics, University of Warwick, Gibbet Hill Road, Coventry, UK \\
$^{3}$Space Telescope Science Institute, 3700 San Martin Drive, Baltimore, MD 21218, USA \\
$^{4}$Inter-University Institute for Data Intensive Astronomy, Department of Astronomy, University of Cape Town, Private Bag X3, Rondebosch 7701, South Africa \\
$^{5}$South African Astronomical Observatory, P.O. Box 9, 7935 Observatory, South Africa \\
$^{6}$Department of Physics, The George Washington University, Corcoran Hall, 725 21st St NW, Washington, DC 20052, USA \\
$^{7}$GWU/Astronomy, Physics and Statistics Institute of Sciences (APSIS) \\
$^{8}$School of Physics and Astronomy, University of Leicester, University Road, Leicester, LE1 7RH, UK \\
$^{9}$Physics Department, Lancaster University, Lancaster, LA1 4YB, UK \\
}

\date{Accepted XXX. Received YYY; in original form ZZZ}

\pubyear{2022}

\begin{document}
\label{firstpage}
\pagerange{\pageref{firstpage}--\pageref{lastpage}}
\maketitle

\begin{abstract}
We report the discovery of six new magnetar counterpart candidates from deep near-infrared Hubble Space Telescope imaging. The new candidates are among a sample of nineteen magnetars for which we present HST data obtained between 2018--2020. We confirm the variability of previously established near-infrared counterparts, and newly identify candidates for PSR\,J1622-4950, Swift\,J1822.3-1606, CXOU\,J171405.7-381031, Swift\,J1833-0832, Swift\,J1834.9-0846 and AX\,J1818.8-1559 based on their proximity to X-ray localisations. The new candidates are compared with the existing counterpart population in terms of their colours, magnitudes, and near-infrared to X-ray spectral indices. We find two candidates for AX\,J1818 which are both consistent with previously established counterparts. The other new candidates are likely to be chance alignments, or otherwise have a different origin for their near-infrared emission not previously seen in magnetar counterparts. Further observations and studies of these candidates are needed to firmly establish their nature.
\end{abstract}

\begin{keywords}
stars: magnetars
\end{keywords}



\section{Introduction}
Magnetars are neutron stars with extremely high magnetic field strengths ($B\sim10^{14}$\,G) and are of interest in a wide range of astrophysical research areas \citep[see][for a review]{2017ARA&A..55..261K}. They provide natural laboratories to test quantum mechanics and general relativity at the extremes, have the potential to reveal much about the formation of neutron stars in general, and have been invoked as the central engines in transients ranging from gamma-ray bursts to superluminous supernovae \citep[e.g.][]{2014MNRAS.438..240G,2015MNRAS.454.3311M,2017ApJ...841...14M} and fast blue optical transients \citep{2018ApJ...865L...3P,2020ApJ...888L..24M}. Recently, they have also been suggested as promising candidates for the origin of extragalactic fast radio bursts (FRBs), a theory backed up by the detection of low-luminosity FRB-like bursts from Galactic magnetar SGR\,1935$+$2154 \citep{2020Natur.587...54C,2020Natur.587...59B}.

A more direct way to study magnetars is to measure the multi-wavelength emission from Galactic sources, of which $\sim$30 are known \citep{2014ApJS..212....6O}\footnote{\url{http://www.physics.mcgill.ca/~pulsar/magnetar/main}}. Magnetar emission is distinct from the magnetic dipole braking radiation seen from pulsars - the persistent magnetospheric emission is likely driven by the direct decay of the intense magnetic field \citep{2002ApJ...574..332T}, with bursts and flares driven by magnetic reconnections \citep[akin to solar flares,][]{2003MNRAS.346..540L}, starquakes \citep{1995MNRAS.275..255T,2007ApJ...657..967B} or other electrodynamic processes \citep{2005ApJ...618..463H}. Galactic magnetars are typically discovered through $\gamma$-ray/hard X-ray flares, with confirmation following the detection of a coincident persistent X-ray source \citep[or in some cases a radio source,][]{2017ARA&A..55..261K}. A population of magnetars has now been identified outside the Milky Way, through the detection of giant flares \citep{2021ApJ...907L..28B}.

Other than at ${\gamma}$-ray, X-ray and radio wavelengths, a few magnetars have also been observed in the optical and infrared \citep[e.g.][]{2004A&A...416.1037H,2005ApJ...623L.125K,2007ApJ...669..561C,2008A&A...482..607T,2011MNRAS.416L..16D,2012ApJ...761...76T}. Given their locations in the Galactic plane, high dust extinctions have restricted these observations, but optical/near-infrared (NIR) variability has been noted in a handful of cases. Sometimes this variability is correlated with X-ray activity \citep{2004ApJ...617L..53T,2006ApJ...640..435E,2008ApJ...677..503T}, which can be explained through the presence of a debris disc, formed through supernova fallback, and heated by emission from the magnetar \citep{2006Natur.440..772W,2006ApJ...649L..87E,2014ApJ...796...46O,2016ApJ...833..265T}. Magnetospheric models for the origin of the NIR counterpart also predict a correlation with X-ray emission \citep{2007ApJ...657..967B}. However, the NIR and X-rays do not always vary synchronously \citep[][]{2007ApJ...669..561C,2008A&A...482..607T,2021arXiv211207023L}. The aforementioned variability has been reported for a handful of sources over month to year time-scales, but short time-scale (${\sim}$seconds) pulsations have also been seen \citep{kern2002,2011MNRAS.416L..16D}. 

Studies thus far have been limited by to the small optical/NIR counterpart population size, which has made it difficult to determine whether all magnetars have similar emission properties at these wavelengths. In this paper, we present Hubble Space Telescope ({\em HST}) imaging of 19 Galactic magnetars. We perform photometry on previously suggested counterparts and 6 newly identified candidates reported here for the first time. We compare the magnitudes of the previously known sources to this latest epoch of observations, confirming or establishing NIR variability, and compare the properties of the new candidate counterparts to the existing population. 

This paper is structured as follows. We describe the details of the observations and counterpart candidate identification in Section \ref{sec:cp}, perform photometry in Section \ref{sec:phot}, and plot the sources on colour-magnitude diagrams in Section \ref{sec:colours}. The variability and spectral indices of the candidates are discussed in Sections \ref{sec:var} and \ref{sec:index}, followed by conclusions in Section \ref{sec:conc}. Throughout, magnitudes are reported in the Vega system; appropriate conversions from AB magnitudes \citep{1983ApJ...266..713O} have been applied where necessary \citep[either following][or with stsynphot\footnote{\url{https://github.com/spacetelescope/stsynphot_refactor}} for \hst\ filters]{2007AJ....133..734B}. The additions to \hst\ Vega magnitudes to obtain AB magnitudes are 0.9204 (F125W), 1.0973 (F140W) and 1.2741 (F160W).

\section{Observations and Counterpart identification}\label{sec:cp}

\subsection{{\em HST} observations}
Details of the 19 datasets used in this work are listed in Table \ref{tab:data}, along with abbreviated magnetar names which we use throughout. All except CXOU\,J1647 have F125W ($\sim$J-band) and F160W ($\sim$H-band) imaging, CXOU\,J1647 has just F140W. All observations were taken with WFC3/IR. There are 20 datasets listed, but guide star acquisition failed for SGR\,1806. The previously unpublished images, obtained from the archive, were corrected for charge transfer efficiency (CTE) and reduced with standard {\sc drizzlepac} procedures \citep[with default settings,][]{2021AAS...23821602H}. The native pixel scale was maintained, i.e. {\sc pixfrac}$=$1 and final scale 0.1265\,arcsec\,pixel$^{-1}$. Another source with recent \hst\ observations is SGR\,1935. We have not included it here as the data is already published and discussed in detail by \citet{2018ApJ...854..161L} and \citet{2021arXiv211207023L}.

\begin{table}
\centering 
\caption{Details of the \hst\ data. Shortened magnetar names (without brackets) are used throughout the paper. The data is primarily from programmes 15348 and 16019 (PI: Levan). The listed exposure times are repeated, once in F160W and once in F125W in each case, with the exception of CXOU\,J1647 (prog. 14805), for which an F140W exposure was performed twice. WFC3 was used for all observations. }
\label{tab:data}
\begin{tabular}{p{3.2cm}p{1.1cm}p{1.8cm}p{0.8cm}}
\hline %
Magnetar  & Prog. & Date & Exp [s] \\
\hline %
4U\,0142($+$61) &	15348	&	2018 Jan 01	&		598 \\
SGR\,0418($+$5729)	&	16019	&	2020 Jan 29	&		898 \\
SGR\,0501($+$4516)	&	16019	&	2020 Aug 04	&		598 \\
1E\,1547(.0-5408)	&	15348	&	2018 Sep 03	&		898 \\
PSR\,J1622(-4950)	&	16019	&	2020 Sep 04	&		898 \\
SGR\,1627(-41)	&	16019	&	2020 Sep 08	&		898 \\
1RXS\,J1708(49.0-400910)	&	15348	&	2018 Oct 05	&		898 \\
CXOU\,J1714(05.7-381031)	&	15348	&	2018 Apr 13	&		898 \\
SGR\,1745(-2900)	&	15348	&	2018 Apr 09	&		898 \\
SGR\,1806(-20)\,${\dagger}$	&	15348	&	2018 Jun 18	&		598 \\
XTE\,J1810(-197)	&	15348	&	2018 Aug 03	&		898 \\
Swift\,J1822(.3-1606)	&	15348	&	2018 Jul 05	&		598 \\
SGR\,1833(-0832)	&	15348	&	2018 Jul 18	&		898 \\
Swift\,J1834(.9-0846)	&	16019	&	2020 Mar 16	&		898 \\
3XMM\,J1852(46.6$+$003317)	&	15348	&	2018 Jun 14	&		898 \\
SGR\,1900($+$14)	&	16019	&	2020 Aug 26	&		898 \\
1E\,2259($+$586)	&	15348	&	2018 Aug 16	&		598 \\
SGR\,0755(-2933)	&	16019	&	2020 Sep 10	&		898 \\
AX\,J1818(.8-1559)	&	15348	&	2018 Aug 02	&		898 \\
CXOU\,J1647(10.2-455216)	&	14805	&	2018 May 23	&		2688 \\
\hline
\hline %
\end{tabular}
\newline
${\dagger}$ - guide star not acquired, images unusable \\
\end{table}

\subsection{Counterpart localisation with {\em Chandra}}
\begin{figure*}
    \centering
	\includegraphics[width=0.99\textwidth]{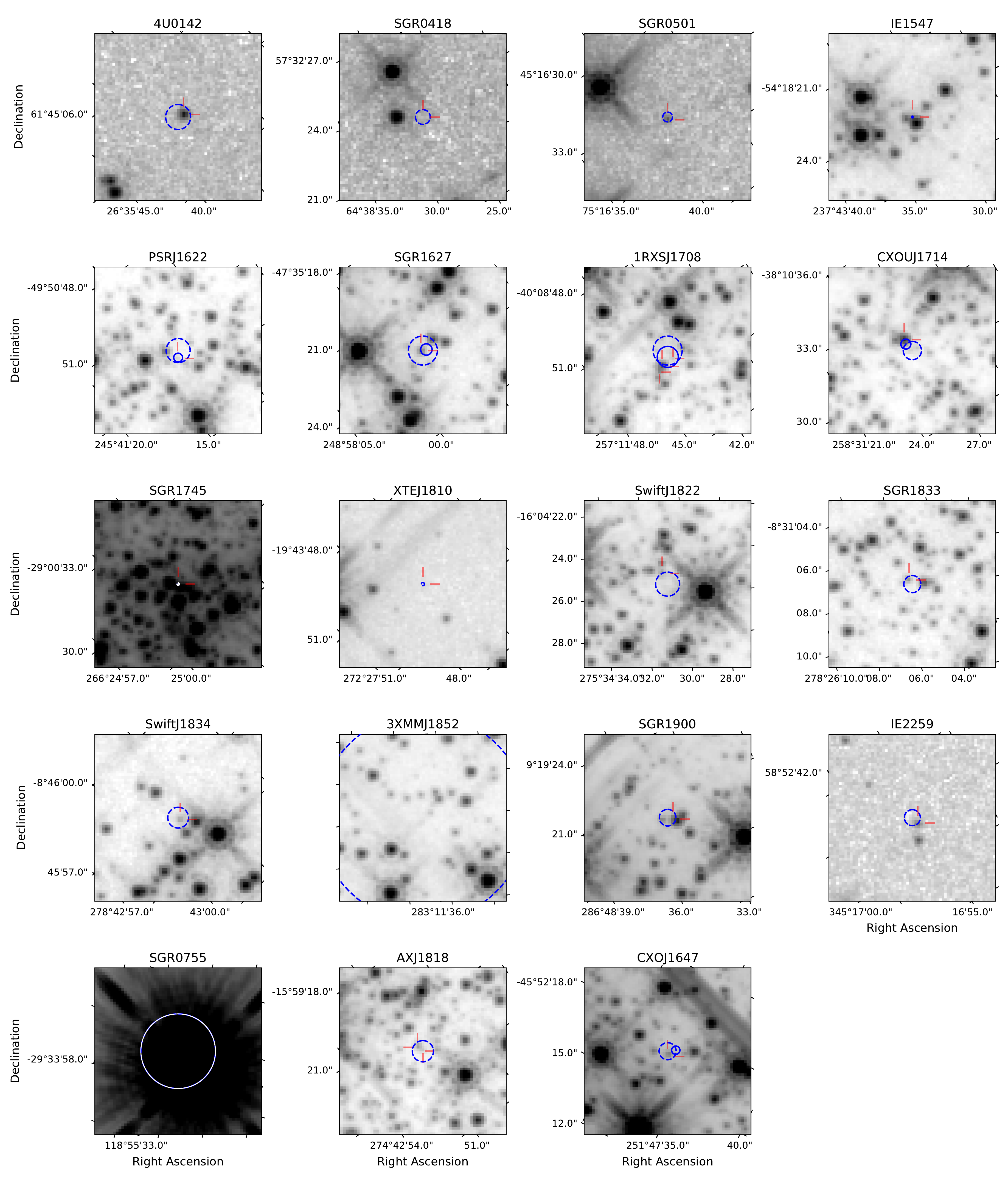}
    \caption{F160W \hst\ imaging of the 19 magnetars (with the exception of CXOU\,J1647 which has F140W imaging). The cutouts are 8${\times}$8 arcsec. Error circles which have been refined through relative alignment (e.g. \cxo\ with \hst\ ) are drawn as solid lines. Error circles based on absolute astrometry are dashed (if this uncertainty is lower than from relative astrometry, we only show the absolute localisation). In each case the error circle radii are adjusted to enclose $\sim$95 per cent of the probability. If sources are detected (or have previously been detected) in the error circle, the most likely counterpart position(s) is indicated by red pointers. This location is used to measure a limit in the case of non-detections. If there is no previously reported counterpart, the aperture is placed at the centre the X-ray error circle. There are three candidates for 1RXSJ\,1708, including one outside the error circle which we measure for comparison with \citet{2006ApJ...648..534D}. AXJ\,1818 has two candidates. The error circle for 3XMM\,J1852 is larger than the cutout. For SGRs\,0755 and 1745, the error circle is drawn in white for contrast. The corresponding photometry is listed in Table \ref{tab:samples}.}
    \label{fig:stamps160}
\end{figure*}

\begin{table*}
\caption{Magnetar positions reported in the literature, and details of the \cxo -\hst\ alignment described in the text. Where the number of tie sources, $N_{\rm tie}$, is listed as X$/$Y, the first number is the objects matched between \cxo\ and the intermediate, and the second between the intermediate and \hst . Refined 2$\sigma$ uncertainty radii from relative astrometry are listed in the final column, these are used if they provide an improvement over the absolute astrometry, and 2 or more tie objects were available. Where we have been unable to precisely align a \cxo\ X-ray localisation on the \hst\ images, we have re-calculated the absolute astrometric uncertainty in these cases using the net source counts from {\sc wavdetect} and following \citet{2010ApJS..189...37E}. These uncertainties are denoted with a $\dagger$ symbol.} 
\label{tab:loc}
\begin{tabular}{lllllllll}
\hline %
Magnetar	&	Localisation/identification	&	RA	&	Dec	&	Unc. (enc.)	&	CXO	&	$N_{\rm tie}$	&	Intermediate	&	2${\sigma}$ unc.	\\
	&	reference	&	(as reported)	&	(as reported)	&	[arcsec]	&	Obs ID	& 	&		&	[arcsec]	\\
\hline																	
4U\,0142	&	\protect{\cite{2002ApJ...568L..31J}}	&	01h46m22.41s	&	+61d45m03.2s	&	0.60 (95)$\dagger$	&	723	&	1 / 61	&	\gaia\ 	&	-	\\
SGR\,0418	&	\protect{\cite{2010ApJ...711L...1V}}	&	04h18m33.867s	&	+57d32m22.91s	&	0.35 (95)	&	10168	&	 0	&	\gaia\ 	&	 -	\\
SGR\,0501	&	\protect{\cite{2010ApJ...722..899G}}	&	05h01m06.76s	&	+45d16m33.92s	&	0.11 (67)	&	9131	&	 0	&	\gaia\ 	&	 -	\\
1E\,1547	&	\protect{\cite{2012ApJ...748L...1D}}	&	15h50m54.12s	&	-54d18m24.1s	&	(0.6$\times$2)$\times$10$^{-3}$ (67)	&	 {\it radio}	&	 -	&	 -	&	 -	\\
PSR\,J1622	&	\protect{\cite{2012ApJ...751...53A}}	&	16h22m44.89s	&	-49d50m52.7s	&	0.58 (95)$\dagger$	&	10929	&	2	&	None	&	0.210	\\
SGR\,1627	&	\protect{\cite{2004ApJ...615..887W}}	&	16h35m51.844s	&	-47d35m23.31s	&	0.74 (95)$\dagger$	&	1981	&	3	&	None	&	0.272	\\
1RXS\,J1708	&	\protect{\cite{2003ApJ...589L..93I}}	&	17h08m46.87s	&	-40d08m52.4s	&	0.7 (95)$\dagger$	&	1936	&	2	&	None	&	0.502	\\
CXOU\,J1714	&	\protect{\cite{2010ApJ...725.1384H}}	&	17h14m05.74s	&	-38d10m30.9s	&	0.44 (95)$\dagger$	&	6692	&	2	&	None	&	0.238	\\
SGR\,J1745	&	\protect{\cite{2013MNRAS.435L..29S}}	&	17h45m40.16s	&	-29d00m29.8s	&	(9$\times$22)$\times$10$^{-3}$ (67)	&	 {\it radio}	&	 -	&	 -	&	 -	\\
XTE\,J1810	&	\protect{\cite{2007ApJ...662.1198H}}	&	18h09m51.087s	&	-19d43m51.93s	&	4$\times$10$^{-3}$ (67)	&	 {\it radio}	&	 -	&	 -	&	 -	\\
Swift\,J1822	&	\protect{\cite{2012ApJ...761...66S}}	&	18h22m18.06s	&	-16d04m25.5s	&	0.58 (95)$\dagger$	&	15992	&	 0	&	 \gaia\ 	&	 -	\\
SGR\,1833	&	\protect{\cite{2010ApJ...718..331G}}	&	18h33m44.37s	&	-08d31m07.5s	&	0.41 (95)$\dagger$	&	11114	&	2 / 85	&	\gaia\	&	0.588	\\
3XMM\,J1852	&	\protect{\cite{2014ApJ...781L..16Z}}	&	18h52m46.67s	&	+00d33m17.8s	&	2.4 (67)	&	{\it no CXO}	&	 -	&	 -	&	 -	\\
SGR\,1900	&	\protect{\cite{1999Natur.398..127F}}	&	19h07m14.33s	&	+09d19m20.1s	&	0.4 (95)$\dagger$	&	6731	&	2 / 42	&	Pan-STARRS	&	0.494	\\
1E\,2259	&	\protect{\cite{2001ApJ...563L..49H}}	&	23h01m08.295s	&	+58d52m44.45s	&	0.38 (95)$\dagger$	&	6730	&	2 / 101	&	\gaia\	&	0.41	\\
SGR\,0755	&	\protect{\cite{2021A&A...647A.165D}}	&	07h55m42.48s	&	-29d33m49.2s 	&	2.0 (97)	&	22454	&	1 / 116	&	\gaia\ 	&	-	\\
Swift\,J1834	&	\protect{\cite{2012ApJ...748...26K}}	&	18h34m52.118s	&	-08d45m56.02s	&	0.5 (95)$\dagger$	&	 14329	&	 0	&	\gaia\ 	&	 -	\\
AXJ1818	&	\protect{\cite{2012A&A...546A..30M}}	&	18h18m51.38s	&	-15d59m22.62s	&	0.51 (95)$\dagger$	&	7617	&	 1/148	&	 \gaia\ 	&	 -	\\
CXOU\,J1647	&	\protect{\cite{2006ApJ...636L..41M}}	&	16h47m10.20s	&	-45d52m16.90s	&	0.41 (95)$\dagger$	&	19136	&	2 / 188	&	\gaia\	&	0.198	\\
\hline %
\hline %
\end{tabular}
\end{table*}

{\it Chandra X-ray Observatory} (\cxo) observations provide the best localisation available for 15 of the 19 magnetars. We download the  \cxo\ event files for each source, obtained via the \cxo\ data centre with the Obs IDs as listed in Table \ref{tab:loc}. Observations are variously with ACIS and HRC. We measure the source positions in these images using the CIAO source detection algorithm {\sc wavdetect}. Standard {\sc ciao} (v4.13, with caldb v4.9.3) procedures were used, including reprocessing, PSF map creation and energy filtering to the range 0.5-7\,keV (HRC) or 0.5-8\,kev (ACIS). Point sources 5\,${\sigma}$ above the background level are extracted. 

As a first step, we attempt to find sources in common between the X-ray and \hst\ images, and compute the offsets in RA and Dec required to map one set of coordinates to the other (astrometrically ‘tying’ the images). In this way, we are removing the absolute uncertainty of the world coordinate system (WCS) calibrations and are limited only by the relative uncertainty in the transformation. Ideally we would find sources in common directly between \cxo\ and \hst, which was only possible in 4 cases. The source centroids were measured in each image’s own coordinate system, and the mean RA and Dec offsets needed to map \cxo\ coordinates on to the \hst\ frame were computed. These were applied to the \cxo\ coordinates, and the RMS difference between the transformed \cxo\ and \hst\ source positions measured. This is added in quadrature to the centroid uncertainty, calculated as FWHM/(2.35\,SNR), where FWHM is the full width at half maximum and SNR is the signal to noise ratio. This yields the smallest possible (sub-arcsecond) positional uncertainties. We note that the resulting refined error circle for CXOU\,J1647 is slightly offset from the source identified as the counterpart by \citet{2008A&A...482..607T}. However, as there are no other objects in the unrefined or refined error circle, we adopt this source as the likely counterpart.

For the remaining 11 sources with \hst\ and \cxo\ observations, there are $<2$ sources in common that are not a counterpart candidate, precluding a direct tie between \hst\ and \cxo\ (without relying on a single tie object). We instead try to tie the images via an intermediate step. The intermediate image should have an area comparable to \cxo\ images, so that common sources can be found within the \cxo\ field of view, but deep enough for sources in common with \hst\ to be identified. Intermediate catalogues searched include \gaia\ \citep[EDR3,][]{2021A&A...649A...1G}, 2MASS \citep{2006AJ....131.1163S} and Pan-STARRS \citep{2016arXiv161205560C}. If there were insufficient matches in these surveys, we also searched the ESO and Gemini archives for imaging of the fields. In each case, where \gaia\ offered enough suitable intermediate tie objects, no improvement could be gained by moving to a different survey. The uncertainty in the \cxo\ - intermediate alignment is calculated in the same way as for the \cxo\ - \hst\ direct alignments, by computing the RA and Dec shifts required and the RMS of this translation. For the intermediate - \hst\ alignment, there are tens to hundreds of matches, so in principle rotation and scaling could also be left as free parameters. This is equivalent to improving the absolute astrometry of the \hst\ frame. The intermediate ({\it Gaia}) - \hst\ RMS values, using simple RA and Dec shifts, are typically $\sim$20\,mas (e.g. for 1E\,2259, 25\,mas). This compares well to the RMS values obtained from the FITS headers of \hst\ advanced data product images (for 1E\,2259, 23\,mas in RA and 26\,mas in Dec), whose astrometric solutions are now calibrated against {\it Gaia} (DR1 or DR2) by the automated pipeline. In any case, the \cxo\ - intermediate step dominates the uncertainty in the transformation. Using 1E\,2259 as an example again, the \cxo\ - \gaia\ RMS is 0.2\,arcsec, so any further refinement of the \hst\ - intermediate step will be of little benefit. 

Details of the image alignment and tying are given in Table \ref{tab:loc}. Other datasets are available to act as intermediates for for 4U\,0142 and SGR\,0501, but these counterparts are already well established and unambiguously identified. The refined 2${\sigma}$ error radii have three components if an intermediate is used - the two tie steps and the centroid error. Error circles from relative astrometry are drawn as solid circles in Figure \ref{fig:stamps160}, with candidate counterparts (or their expected location in the case of non-detections) indicated by red pointers. 

Where a \hst\ - \cxo\ alignment was not possible even via an intermediate, error circles are placed on the \hst\ images at the coordinates reported by the references in Table \ref{tab:loc}. The absolute astrometric accuracy of \cxo\ is variously reported in these references as $\sim$0.6-0.8\,arcsec (90 per cent)\footnote{\url{https://cxc.harvard.edu/cal/ASPECT/}}\footnote{\url{https://cxc.harvard.edu/ciao/ahelp/coords.html}}, but the precise value depends on the source location in the image and net counts. We re-calculate the uncertainty in the absolute astrometry of these sources by taking the net counts from {\sc wavdetect} and applying the (off-axis angle and source counts dependent) absolute accuracy calculations of \citet{2010ApJS..189...37E}. Since many of these sources are bright, and the off-axis angles are exclusively $\sim$0 (they are the targets), the absolute astrometric uncertainties are typically smaller than previously reported. The sources for which we have performed this calculation are noted in Table \ref{tab:loc}. If these uncertainties are smaller than those arising from relative alignment, we use the absolute astrometry instead. In one instance an accurately calculated value is already reported (SGR\,0418), and in another, the absolute astrometry of the \cxo\ localisation was improved by alignment with 2MASS (SGR\,0501). 

Two sources have non-\cxo\ localisations plotted in Fig. \ref{fig:stamps160}. These are an XMM position for 3XMM\,J1852 and a {\it Swift}/XRT position of SGR\,0755 - which we plot instead of the later \cxo\ localisation, since this XRT position was associated with the BAT SGR discovery burst. While 3XMM\,J1852 has \cxo\ observations, they are exclusively in continuous clocking mode, so they cannot be used for imaging. The magnetar is also in the field of view of Kes 79 supernova remnant observations, but is not detected in these images.

When placing positions from absolute astrometry on our \hst\ images , we must also consider that the \hst\ image WCS solutions also have an uncertainty. This is quantified by the RA and Dec RMS values after source alignment with \gaia\ as performed by the \hst\ pipeline, of order tens of mas. Assuming the \cxo\ uncertainty is Gaussian, we convert the reported error radius to a 67 per cent radius if it is not already. We add this in quadrature to the \hst -\gaia\ RMS and X-ray centroid uncertainty. In this way we can derive an approximate 2\,${\sigma}$ positional uncertainty for the X-ray source in the \hst\ frame. These error circles are drawn with dashed lines in Figure \ref{fig:stamps160}. Measured proper motions for magnetars are at the mas\,yr$^{-1}$ level, such that temporal separations between the \cxo\ and \hst\ epochs should only produce small spatial offsets compared to the localisation uncertainty.

\subsection{Radio localisations}
The final three sources in our sample have precise radio localisations. The position of 1E\,1547 was measured using very long baseline interferometry observations \citep{2012ApJ...748L...1D}. An ellipsoidal uncertainty region of (0.6${\times}$2.0)\,mas (1\,${\sigma}$) is reported. To place this on the \hst\ frames, however, the uncertainty in the VLBA and \hst\ absolute astrometry must be considered. The quoted VLBA positional error includes both statistical and systematic uncertainties, while the \hst\ frames have been aligned with the \gaia\ reference frame, with an RMS uncertainty in the \hst\ absolute astrometry of 14\,mas (from the RMS$\_$RA and RMS$\_$DEC header entries). The total uncertainty on the magnetar position in the \hst\ frame is, therefore, dominated by the HST-\gaia\ alignment. 

Since SGR\,1745 lies near the Galactic centre, the extremely high extinction along this sightline means that the chance of any sources seen in the crowded field being associated (rather than foreground objects) is very low. Nevertheless, we can measure a limit at the position on the \hst\ image. As with 1E\,1547, we assume the uncertainty on this precise ATCA localisation in the \hst\ frame is dominated by the \hst\ WCS solution \citep{2013MNRAS.435L..29S}. The final magnetar in this sample where we use a radio-localisation is XTE\,J1810. \citet{2007ApJ...662.1198H} report an absolute astrometric uncertainty of 4\,mas, so the HST-\gaia\ WCS solution again dominates.

\subsection{Identification through variability}\label{sec:varloc}

Some NIR counterparts have been identified due to their variability. In these cases, the variable source association is favoured (over non-variable sources in the X-ray error circle) either because the NIR variability correlates with X-ray behaviour, or the expectation that variable NIR sources are sufficiently rare that finding such a source in the error circle by chance is extremely unlikely (although such arguments are as yet poorly quantified).

\begin{figure}
	\centering
	\includegraphics[width=0.9\columnwidth]{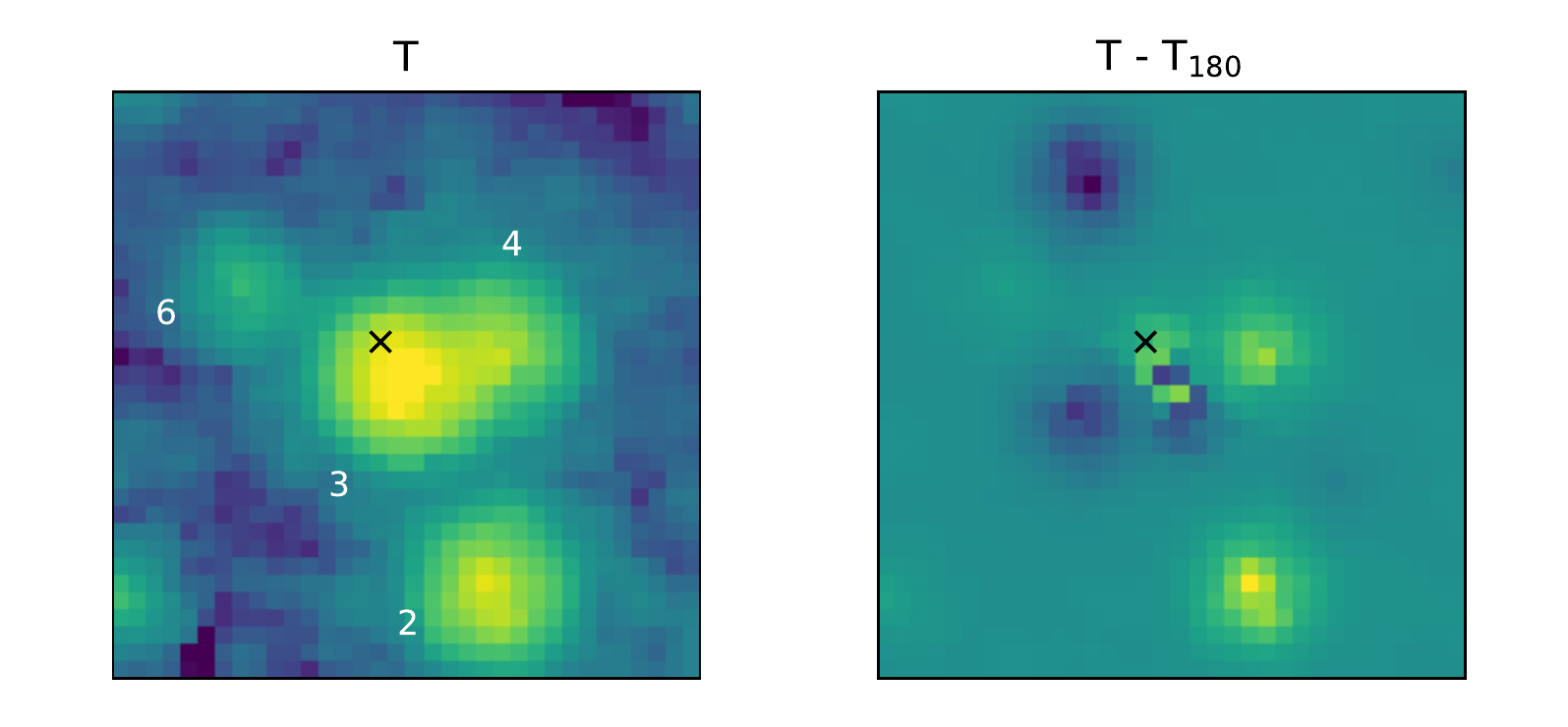}
    \caption{Left: a 2.2$\times$2.2\,arcsec cutout of the re-drizzled F160W image (pixel scale 0.065\,arcsec\,pixel$^{-1}$) centered around the target star (T) coincident with SGR\,1900 \citep[star 3 of][other sources also follow this labelling]{2012ApJ...761...76T}. The approximate location of star 7, the SGR\,1900 counterpart candidate, is marked by a cross. Right: the target cutout rotated by 180 degrees, subtracted from the original cutout. There is a slight excess at the location of sources 7 and 10 in the residual image (mirrored in x and y by an equal deficit since we subtracted a rotated image).}
    \label{fig:1900}
\end{figure}

The source associated with SGR\,1900 has $H = 21.17\pm0.04$ and $K_{p} = 20.63\pm0.02$ (with 0.5 and 0.1 errors on the zero points), and was suggested as the counterpart due to variability \citep{2012ApJ...761...76T}. The two clear sources within/on the edge of the error circle in the \hst\ imaging presented in Figure \ref{fig:stamps160} are labelled as sources 3 and 6 by \citet{2012ApJ...761...76T} \citep[and][]{2008A&A...482..607T}. The suggested variable counterpart, source 7, and another object, source 10, are not readily visible in the \hst\ images, nor are they visible in images re-drizzled onto a 0.065\,arcsec\,pixel$^{-1}$ grid. To investigate whether proximity to the bright star 3 is obscuring sources 7 and 10 at this resolution, we subtract the PSF of this star in Figure \ref{fig:1900}. To do this, we subtract a rotated image cutout centred on star 3, to sample the precise PSF at this location. There is slight excess in the residual image between stars 3 and 6, $\sim$0.2\,arcsec from the centre of star 3, at approximate location of stars 7 and 10 in the Keck/NIRC2 LGS-AO observations of \citet{2012ApJ...761...76T}. The reported proper motion of 2.1$\pm$0.6\,mas\,yr$^{-1}$ for the magnetar corresponds to only $\sim$0.3 pixels of movement over 10 years at this pixel scale. We conclude that we are not resolving these sources from star 3. The high flux ratio (a magnitude $\sim$18 source only 0.2\,arcsec from magnitude $\sim$21 sources) makes robust recovery of the suggested counterpart extremely challenging, so we consequently adopt the range of previously reported $H$ and $K$-band photometry for the suggested variable counterpart \citep{2008A&A...482..607T,2012ApJ...761...76T}.

\citet{2012ApJ...761...76T} also identify a candidate counterpart to SGR\,1806 thanks to its variability, measuring $K\sim21.75\pm0.75$ (this approximate uncertainty reflects the source variability), in agreement with \citet{2005ApJ...623L.125K} who find $K=21.9$. We adopt these values since the \hst\ images for this target are unusable following a guide star acquisition failure. 

There have been unambiguous localisation (both in previous works and here) of magnetar counterparts known to be variable, for example 1E\,2259 and 4U\,0142 \citep{2004ApJ...617L..53T,2004A&A...416.1037H,2013ApJ...772...31T}. Since they are clearly localised with \cxo\, independent of variability arguments, this demonstrates that at least some magnetar counterparts are variable and supports associations based on variability.

\subsection{New counterpart candidates}
Using the localisations and imaging described above, we identify counterpart candidates for six magnetars for the first time. These are PSR\,J1622, Swift\,J1822, CXOU\,J1714, Swift\,J1833, Swift\,J1834 and AX\,J1818 (two candidates). This has increased the sample of Galactic magnetar NIR counterparts (confirmed or otherwise) by $\sim$50 per cent \citep{2014ApJS..212....6O}. Additionally, SGR\,1627 has previous imaging, but the source we deem to be the most likely counterpart has not been measured, as it was too faint for reliable photometry in previous imaging \citep{2004ApJ...615..887W}. We therefore report photometry for this candidate for the first time.

\begin{figure*}
	\includegraphics[width=0.9\textwidth]{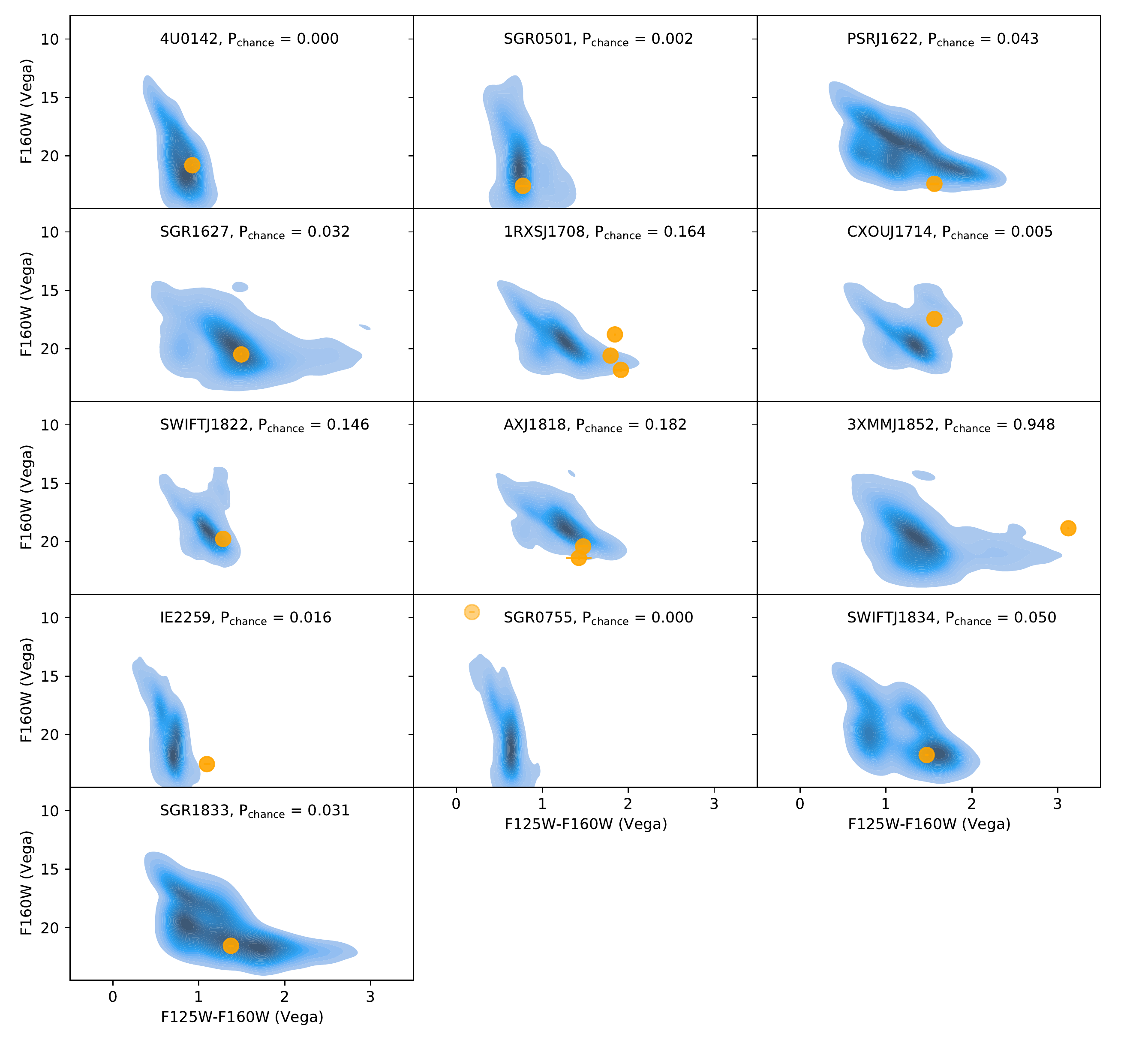}
    \caption{Colour-magnitude diagrams of the fields of the 13 magnetars with dual-band imaging and a detection. The source photometry was extracted in both filters simultaneously using DOLPHOT. The density of field objects in this parameter space is indicated by the blue shading, targets in the error circles in Figure \ref{fig:stamps160} are shown as orange points (errorbars are typically too small to be visible). Unusually red colours, outside the cloud of stellar sources, might indicate a non-stellar origin (e.g. a debris disc), however this is not necessarily the case: the spectral energy distribution of 4U\,0142 suggests the existence of a debris disc, but it does not appear particularly anomalous in this parameter space. The probability of finding a source in the error circle, of the counterpart brightness or brighter, is given by P$_{\rm chance}$ (the brightest counterpart candidate is used where there are several). The 3XMM\,J1852 source is simply the reddest object in the large error circle, and is not a robust counterpart candidate.}
    \label{fig:obsvcmd}
\end{figure*}

\section{Photometry}\label{sec:phot}
\subsection{Manual photometry}
We initially perform manual photometry on counterparts or candidates identified in the 2${\sigma}$ error circles as shown in Figure \ref{fig:stamps160} and listed in Table \ref{tab:samples}. Aperture photometry is carried out with the {\sc photutils} python package \citep[v1.2.0,][]{larry_bradley_2021_4624996}. The aperture radius used is typically 0.4\,arsec, but is reduced to as low as 0.1\,arcsec if there are immediately neighbouring sources, to reduce flux contamination. The tabulated WFC3 IR encircled energy corrections are applied\footnote{\url{https://www.stsci.edu/hst/instrumentation/wfc3/data-analysis/photometric-calibration/ir-encircled-energy}}. Because many of these fields are crowded, rather than using an annulus, we estimate the background by placing apertures on `blank’ areas of sky (judged by eye) around the image. The background level is extracted by constructing a pixel value distribution from these apertures and using the sigma clipped median.

To verify this method, we also calculate the background with the {\sc photutils} function {\sc background2D} in less crowded fields. In these cases we get similar results. For example, for the NIR counterpart of 4U\,0142, the difference in photometry between aperture background and background2D estimators produce results that differ by much less than the total photometric uncertainty (a 0.008 difference with an error of 0.02). The photometry for the counterparts and candidates is listed in Table \ref{tab:samples}, with multiple rows where there is more than one candidate. If the flux at the magnetar location fails to reach 3${\sigma}$ significance, a 3${\sigma}$ limit is listed instead.

\subsection{Automated photometry}
In addition to the aperture photometry, we also perform automated source detection and photometry for all sources in the \hst\ images. There are two reasons for this, (i) to verify the manual measurements and (ii) to obtain magnitudes and colours for every source in the field. This way, we can see if the counterpart candidates stand out as having unusual magnitudes or colours, or find sources that do. DOLPHOT is used \citep[v2.0,][]{dolphin2000}\footnote{\url{http://americano.dolphinsim.com/dolphot/}}, running on the F160W and F125W bands simultaneously, with the point spread functions of \citet{anderson2016}. Drizzled F160W images are used as the reference, but photometry is performed on the undrizzled {\sc flt} frames. We retain sources with a SNR of at least 3 in each filter, and reject those with an ellipticity greater than 0.2, helping to remove diffraction spikes and some galaxies.  Differences between these measurements and the manual photometry are typically small, but can arise because DOLPHOT accounts for source blending and the PSF, which can have a significant impact on \hst\ photometry in crowded fields \citep{1998A&AS..127..327S}.

\begin{table*}
\caption{The \hst\ photometry for candidates in the error circles in Figure \ref{fig:stamps160}, supplemented by $H$ (or if unavailable, $K$) band data from the literature, where we cannot reliably measure at previously suggested counterpart location. Non-\hst\ data is used for SGR\,1900 and SGR\,1806 \citep[$H$ and $K$-band,][]{2012ApJ...761...76T}, and SGR\,0755 ($H$-band, 2MASS).
Otherwise, even if we do not detect a previously suggested counterpart, we use the \hst\ photometry as measured below. AXJ\,1818 and 1RXSJ\,1708 have multiple candidates, the first two for 1RXSJ\,1708 follow the same labelling as \citet{2003ApJ...589L..93I} and \citet{2006ApJ...648..534D}. Quiescent unabsorbed X-ray fluxes (2-10\,kev) are also listed with the reference for the original measurement. Several unabsorbed fluxes were calculated from different energy ranges by \citet{2014ApJS..212....6O} and they are provided below where the original reference did not give this measurement. In the final column we give the NIR to X-ray power law index.}
\label{tab:samples}
\begin{tabular}{lllllll}
\hline %
Magnetar	&	F160W	&	F125W	&	P$_{\rm chance}$	&	Unabsorbed F$_{\rm X}$ [2-10 keV]	&	X-ray reference	&	NIR to X-ray	\\
	&	(Vega)	&	(Vega)	&		&	[10$^{-12}$erg\,s$^{-1}$\,cm$^{-2}$]	&		&	PL Index	\\
\hline													
4U\,0142	&	20.80$\pm$0.01	&	 21.72$\pm$0.01	&	0.000	&	67.9	&	\protect{\cite{2007MNRAS.381..293R}}	&	0.98$\pm$0.01	\\
SGR\,0418	&	$>$25.12	&	$>$26.08	&	 -	&	0.0020$^{+0.0014}_{-0.0010}$	&	\cite{2013ApJ...770...65R}	&	$>$0.26	\\
SGR\,0501	&	22.56$^{+0.06}_{-0.07}$	&	23.33$\pm$0.07	&	0.002	&	1.7	&	\protect{\cite{2014MNRAS.438.3291C}}	&	0.75$\pm$0.02	\\
1E\,1547	&	$>$20.22	&	$>$22.45	&	 -	&	0.54	&	\protect{\cite{2011A&A...529A..19B}}	&	$>$0.38	\\
PSR\,J1622	&	22.39$\pm0.05$	&	23.95$\pm0.08$	&	0.043	&	0.045$^{+0.063}_{-0.028}$	&	\protect{\cite{2012ApJ...751...53A}}	&	0.33$\pm$0.11	\\
SGR\,1627	&	20.48$\pm$0.01	&	21.97$\pm0.01$	&	0.032	&	0.25$^{+0.17}_{-0.10}$	&	\protect{\cite{2012ApJ...757...68A}}	&	0.32$\pm$0.06	\\
1RXS\,J1708 (A)	&	18.77${\pm}$0.01	&	20.61${\pm}$0.01	&	0.036	&	24.3	&	\protect{\cite{2007Ap&SS.308..505R}}	&	0.66$\pm$0.01	\\
1RXS\,J1708 (B)	&	20.58${\pm}$0.01	&	22.37$\pm$0.02	&	0.104	&	 -	&	 -	&	0.97$\pm$0.02	\\
1RXS\,J1708 (C)	&	21.80${\pm}$0.08	&	23.71$^{+0.14}_{-0.16}$	&	0.164	&	 -	&	 -	&	0.85$\pm$0.01	\\
CXOU\,J1714	&	17.45$\pm$0.01	&	19.01$\pm$0.01	&	0.005	&	2.68$\pm$0.09	&	\protect{\cite{2010PASJ...62L..33S}}	&	0.28$\pm$0.01	\\
SGR\,J1745	&	$>$15.97	&	$>$18.98	&	 -	&	$<$0.013	&	\protect{\cite{2013ApJ...770L..23M}}	&	 -	\\
SGR\,1806 $[1]$	&	21.75$\pm$0.75	&	 -	&	 -	&	18$\pm$1	&	\protect{\cite{2007A&A...476..321E}}	&	0.99$\pm$0.05	\\
XTE\,J1810	&	$>$24.79	&	$>$25.58	&	 -	&	0.029	&	\protect{\cite{2004ApJ...605..368G}}	&	$>$0.53	\\
Swift\,J1822	&	19.76$\pm$0.01	&	21.04$\pm0.02$	&	0.146	&	$<$0.0013	&	\protect{\cite{2012ApJ...761...66S}}	&	$<$-0.34	\\
AXJ1818 (A)	&	21.38$^{+0.14}_{-0.16}$	&	22.80$^{+0.14}_{-0.16}$	&	0.242	&	1.68$^{+0.15}_{-0.16}$	&	\protect{\cite{2012A&A...546A..30M}}	&	0.63$\pm$0.02	\\
AXJ1818 (B)	&	20.40$\pm0.06$	&	21.87$^{+0.06}_{-0.07}$	&	0.182	&	 -	&	 -	&	0.53$\pm$0.02	\\
3XMM\,J1852${\dagger}$	&	18.85$\pm$0.01	&	21.97$\pm$0.02	&	0.948	&	$<$0.001	&	\protect{\cite{2014ApJ...781L..17R}}	&	$<$-0.47	\\
SGR\,1900 $[2]$	&	21.17$\pm$0.50	&	 -	&	 -	&	4.8$\pm$0.2	&	\protect{\cite{2006ApJ...653.1423M}}	&	0.74$\pm$0.04	\\
IE\,2259	&	22.52$\pm0.04$	&	23.61${\pm}$0.05	&	0.016	&	14.1$\pm$0.3	&	\protect{\cite{2008ApJ...686..520Z}}	&	0.99$\pm$0.01	\\
SGR\,0755$[3]$	&	9.52$\pm$0.01	&	9.69$\pm$0.01	&	0.000	&	$<$5.5${\ddagger}$	&	\protect{\cite{2016ATel.8868....1A}}	&	$<$-0.31	\\
Swift\,J1834	&	21.74$\pm0.02$	&	23.21$\pm0.03$	&	0.050	&	$<$0.004	&	\protect{\cite{2012ApJ...757...39Y}}	&	$<$-0.01	\\
SGR\,1833	&	21.56$\pm0.03$	&	22.93$\pm0.02$	&	0.031	&	$<$0.2	&	\protect{\cite{2011MNRAS.416..205E}}	&	$<$0.41	\\
CXOU\,J1647	&	22.20$^{+0.09}_{-0.10}$ (F140W)	&	 -	&	0.095	&	0.25$\pm$0.04	&	\protect{\cite{2013ApJ...763...82A}}	&	0.48$\pm$0.03	\\
\hline %
\hline %
\end{tabular}
\newline
${\dagger}$ - Reddest source in error circle, not a robust counterpart association. ${\ddagger}$ - 0.5--10\,keV \\
$[1]$ - $K$-band photometry from \citet{2012ApJ...761...76T}, $[2]$ - $H$-band from \citet{2012ApJ...761...76T}, includes zero-point uncertainty, see Section \ref{sec:varloc} \\
$[3]$ - HMXB \citep{2021A&A...647A.165D}, listed $H$ and $J$-band photometry from 2MASS.\\
\end{table*}

\section{Colour-magnitude diagrams}\label{sec:colours}
In Figure \ref{fig:obsvcmd} we show colour-magnitude diagrams for the fields where at least one candidate counterpart is detected in the error circle. For each field, the probability of finding an object of magnitude $m$ or brighter in the error circle area is evaluated and shown (for the brightest, lowest probability source if there are multiple candidates) in Figure \ref{fig:obsvcmd}. This is given by 
\begin{equation}
    P_{\rm chance} = 1 - {\rm exp}(-{\pi}r^{2}/{\Sigma}),
\end{equation}
where ${\Sigma}$ is the surface density of sources with F160W magnitude $m$ or brighter, averaged over the \hst\ frame, and $r$ is the 2${\sigma}$ uncertainty radius.

Searching for unusually red magnetar counterparts is well established, following the discovery that some do have such colours \citep{2002ApJ...579L..33W,2003ApJ...589L..93I,2004ApJ...615..887W,2011ApJ...742...77D}. An infrared excess was confirmed in magnetar 4U\,0142, possibly due to a debris disc \citep{2006Natur.440..772W}. In other magnetars the red colour may be magnetospheric \citep{1996ApJ...473..322T}. In either case, these sources are non-stellar in colour, appearing distinct from the cloud of other objects. In Figure \ref{fig:obsvcmd}, most of the counterparts and candidates are not obviously distinct. Even some previously established counterparts are not clearly separated in this $J$-$H$ versus $H$ parameter space \citep[e.g. 4U\,0142,][]{2006Natur.440..772W}. The magnetars which do have unusual sources in their error circle are CXOU\,J1714 (off the main sequence), 1RXS\,1708 \citep[source A of][]{2003ApJ...589L..93I}, 1E\,2259 \citep[previously noted as perhaps being magnetospheric in origin,][]{2001ApJ...563L..49H}, SGR\,0755 and 3XMM\,J1852.

The latter two deserve special attention. For 3XMM\,J1852 the {\it XMM} error circle is nearly 5\,arcsec in diameter. There is one exceptionally red source in the error circle, which we adopt as the potential counterpart going forward, with the caveat that there is no way to firmly establish this association with existing datasets.

SGR\,0755 has a {\em Swift}/XRT localisation (indicated on Figure \ref{fig:stamps160} as this was associated with the BAT discovery burst), and further X-ray observations including a \cxo\ localisation \citep{2021A&A...647A.165D}, which is also aligned with the bright star. The {\em Swift}/XRT error circle is slightly offset from the centre of the bright star, which cannot be explained by the {\it Gaia} proper motion of $\sim$4\,mas\,yr$^{-1}$, given that the gap between {\it Swift} and \hst\ observations is only 4 years. In any case, the probability of chance alignment is negligible, and subsequent X-ray observations have also suggested association with the $H{\sim}9$ Be-star, which has been identified as a high-mass X-ray binary \citep[HMXB,][]{2021A&A...647A.165D}. The source is too bright for \hst\ photometry, so the adopted magnitudes are from 2MASS. The SGR classification was initially due to the {\em Swift}/BAT detection of a magnetar-like burst. The HMXB was seen in subsequent {\em Swift}/XRT and \cxo\ observations, but is not the sole X-ray source in the ${\sim}$3\,arcmin BAT error circle, leaving open the possibility of a chance alignment. The 90 percent uncertainty radius on this localisation of ${\sim}$3\,arcmin is larger than the ${\sim}$2\,arcmin \hst\ field of view. Although we identify several sources red-wards of the main sequence, there are too many to suggest counterpart candidates (and the magnetar, if not in the HMXB itself, could still lie outside of the field of view).

\begin{figure}
    \centering
	\includegraphics[width=0.99\columnwidth]{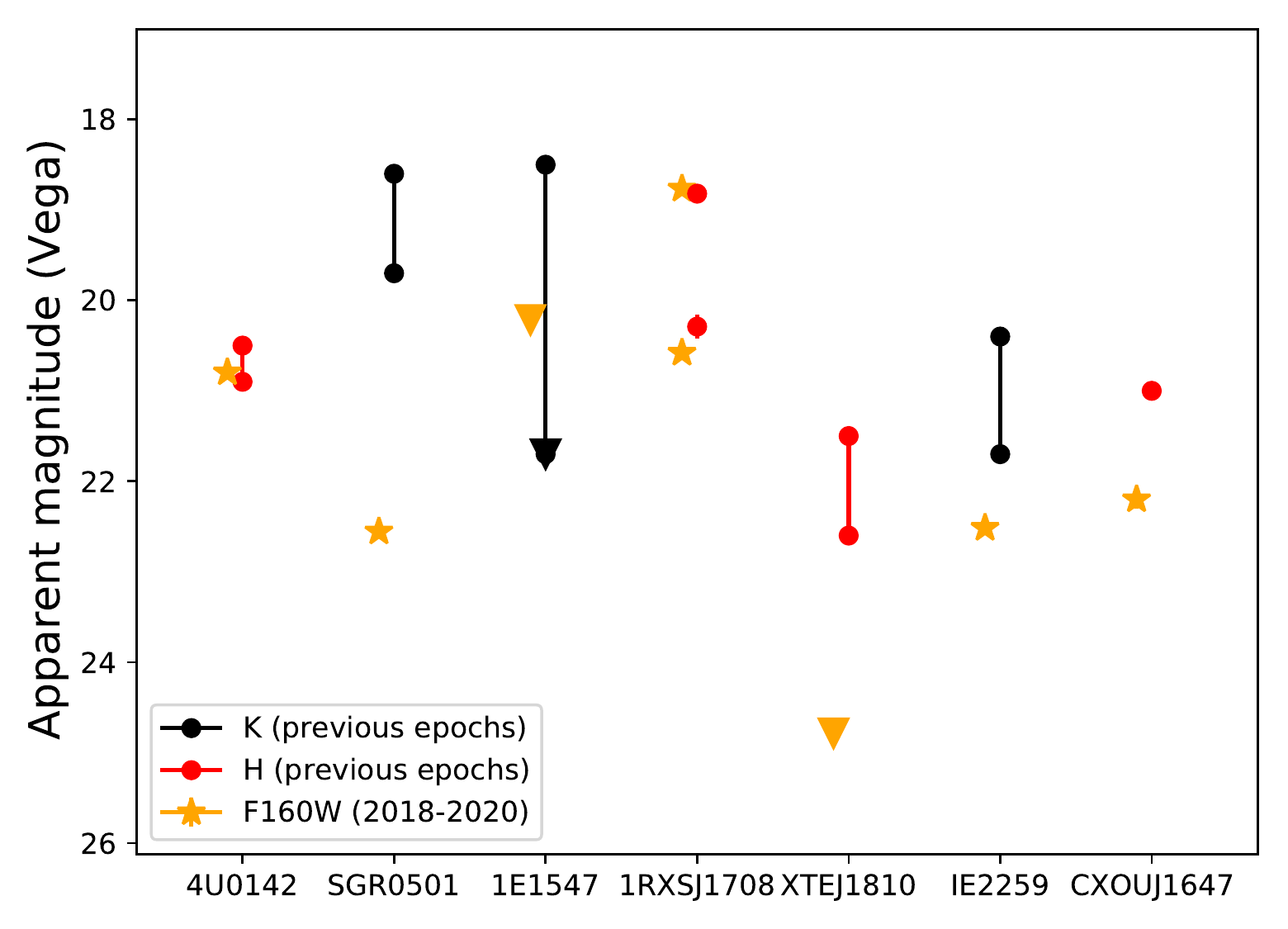}
    \caption{The seven magnetars for which we identify (or place limits on) a previously known NIR counterpart \citep[excluding SGR\,1935, see][for a discussion of the variability of this source]{2021arXiv211207023L}. Shown are the F160W measurements (this work, note that CXOU\,J1647 has F140W observations instead), compared with $H$-band, or if unavailable, $K$-band data from previous studies.}
    \label{fig:variability}
\end{figure}

\section{Variability}\label{sec:var}
In Section \ref{sec:varloc}, we discussed the variability of sources in previous observations as an extra tool for identifying counterparts. Counterpart variability is also interesting in its own right, as it can provide clues to the nature of the emission. In Figure \ref{fig:variability} we show the F160W measurements from this paper, compared to previous measurements of the counterpart ($H$ or $K$-band), where a detection has previously been made \citep[following references in the McGill catalogue,][]{2014ApJS..212....6O}. We also compare with the candidate detection for CXOU\,J1647 by \citet{2018MNRAS.473.3180T}, who find $J=23.5\pm0.2$, $H=21.0\pm0.1$ and $K=20.4\pm0.1$. There are cases where we have detected sources below previously established limits, but these are not constraining in terms of variability and so are not shown. Where there is only $K$-band data to compare, it is difficult to establish if there is variability, or simply a red $H$-$K$ colour (not unusual for magnetars). 

Four have previous $\sim H$-band observations, and a long baseline between observations (4U\,0142, previous observed from 2004--2006, 1RXSJ\,1708, previously observed 2006, XTEJ\,1810, previously observed 2007--2008 and CXOU\,J1647, previously observed 2013). For 1RXSJ\,1708, the second (fainter) source is added from just outside the error circle \citep[labelled star B,][]{2006ApJ...648..534D}. The $H$-band magnitudes of stars A (the brightest object) and B are reported by \citet{2006ApJ...648..534D} to be 18.82$\pm$0.06 and 20.29$\pm$0.13. We report magnitudes of 18.77$\pm$0.01 and 20.58$\pm$0.01, respectively. The star A measurements are therefore consistent, while star B has a ${\sim}$2.2$\sigma$ difference. Given the proximity of star A to star B, and the difference between the $H$ and F160W filters, this is unlikely to be a reliable variability measurement. None the less, \citet{2006ApJ...648..534D} claim the potentially variable star B as the more likely counterpart based on its anomalous $JHK$ colours.

\begin{figure}
    \centering
	\includegraphics[width=0.99\columnwidth]{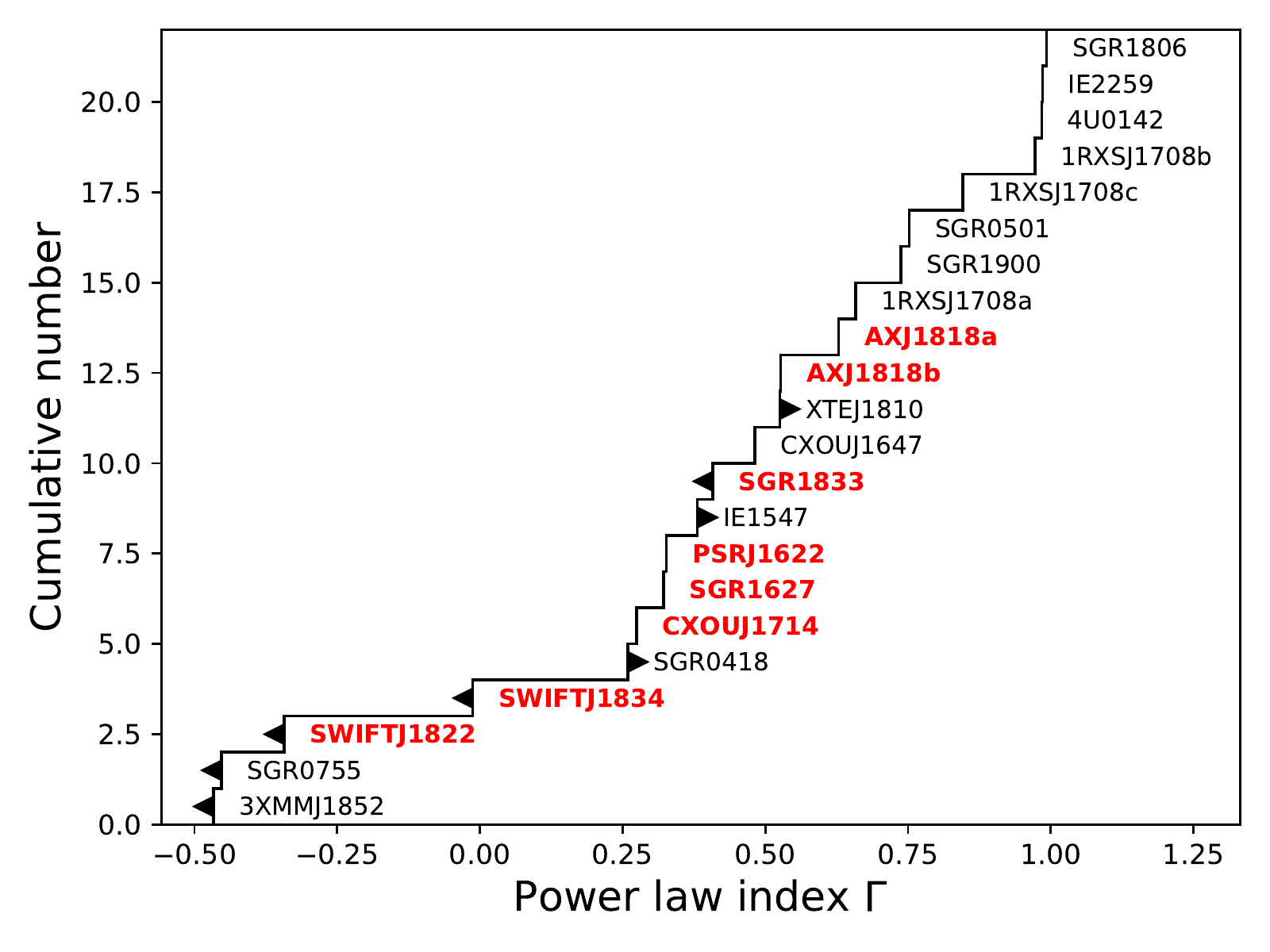}
    \caption{The cumulative distribution of NIR to X-ray power law indices (Table \ref{tab:samples}). Sources for which we report a counterpart candidate for the first time are labelled in bold/red, limits are indicated on the corresponding histogram step. Photon index upper limits are due to X-ray upper limits, and lower limits are NIR non-detections. Only SGR\,1745 (not plotted) lacks a quiescent detection in both the NIR and X-ray and therefore has an unconstrained $\Gamma$. The new candidates have lower indices than previously established counterparts, with the exception of AX\,J1818.}
    \label{fig:nirxray}
\end{figure}

4U\,0142 is consistent with previously noted variability \citep{2004A&A...416.1037H,2006ApJ...652..576D}, and there are significant differences between our measurements of SGR\,0501 and 1E\,2259, although these are also in different bands. CXOU\,J1647 appears to have faded by $\sim$1 magnitude, but the most striking change is in XTEJ\,1810 which has dropped ${\sim}$2.5 magnitudes (a factor of ${\sim}$10 in flux). This is much more than the previously reported $H$-band variability of this source \citep[$\sim$1 magnitude,][]{2008A&A...482..607T}. If the NIR emission is thermal in nature (flux\, $\propto T^{4}$), as in the case of blackbody emission from a debris disc, this corresponds to cooling by a factor of $\sim$2 over a 10 year time-scale. 

Variability of NIR emission can equally be explained if we invoke a surviving binary companion as the origin of the emission. The variability of stars in the NIR is poorly understood \citep[e.g.][]{2020ApJ...891L..37L}, but such a scenario would predict a lack of correlation in the variability seen for different spectral bands in some magnetars. Roughly equal-mass binaries would be likely to place OB-stars as the companions, which can have prominent variability in the NIR \citep{bonanos2009,roquette2020}. Another possibility, given the young age estimates for magnetars, are pre-main sequence companions. These could feasibly remain pre-main sequence for longer than the lifetime of a massive star magnetar progenitor, and are known to be variable in the NIR \citep[e.g.][]{carpenter2001, eiroa2002, alves2008}. We will investigate these possibilities further in a separate paper (Chrimes et al., in prep).

\section{NIR -- X-ray spectral indicies}\label{sec:index}
Another measurement we can make for the sample, placing the new candidate counterparts in context, is the NIR to X-ray spectral index. We fit a power law (PL) between the unabsorbed quiescent X-ray flux, adjusted so that all measurements are in the $\sim$2-10\,keV range \citep[taken from][where necessary]{2014ApJS..212....6O}, and F160W (or nearest available band, see Table \ref{tab:samples}). The corresponding frequencies used are 9$\times$10$^{17}$\,Hz and 2$\times$10$^{14}$\,Hz, with $\nu F_{\nu} \propto \nu^{1-\beta}$ and $\beta = 1 - \Gamma$. The uncertainties are not quantified for all of the X-ray flux estimates, where they are not, a 10 per cent uncertainty is assumed, with the caveat that this may be an underestimate.

We first note that the six sources reported here for the first time have typical apparent magnitudes compared the existing population, with $17.5 < H < 22.5$ (see Table \ref{tab:samples}). Most notable is SGR\,0755, which is by far the brightest source at $H\sim10$. However, there is ambiguity as to whether this Be-star HMXB is associated with SGR\,0755, in which case the accretor would be the magnetar, or whether this is an unrelated HMXB which happened to lie in the 3\,arcmin error region of the {\it Swift}/BAT discovery burst \citep{2021A&A...647A.165D}.

Figure \ref{fig:nirxray} shows the cumulative distribution of PL indices $\Gamma$. Previously established counterparts - those for which a single source has previously been noted due to unambiguous localisation or variability - include 4U\,0142, SGR\,1900, SGR\,1806, 1E\,2259, SGR\,0501, CXOU\,J1647 and XTE\,1810. There is a clear delineation between these known counterparts and the new candidates - all known sources lie at the upper end of the distribution. The lowest (CXOU\,J1647) has $\Gamma = 0.43$, the only new candidates to lie above this are for AX\,J1818. 

To measure these indices, we have used unabsorbed X-ray fluxes and observed NIR fluxes, uncorrected for extinction. Magnetars with previously established NIR counterparts have relatively low extinction estimates \citep[e.g. A$_{\rm v} \sim$3.5 for 4U\,0142,][]{2006Natur.440..772W}. The neutral hydrogen column density assumed to calculate the unabsorbed quiescent X-ray flux of 4U\,0142 is $10^{22}$\,cm$^{-2}$ \citep{2007MNRAS.381..293R}. This corresponds to an extinction of A$_{\rm V}\sim5$ according to the  A$_{\rm V}-$N$_{H}$ relation inferred from X-ray scattering halos and supernova remnants \citep{1995A&A...293..889P}. The unabsorbed X-ray estimate is therefore broadly consistent with a small NIR extinction correction of $\sim 0.6$ magnitudes, given A$_{\rm v} \sim$3.5, a Fitzpatrick extinction curve \citep{1999PASP..111...63F} and R$_{\rm v}=3.1$. This further implies that the power law of index of $\sim$1 for 4U\,0142 is close to the intrinsic, unattenuated value. Similar arguments can be made for other established counterparts, such as 1E\,2259 and SGR\,0501. 

Although extinction is generally low at NIR wavelengths, magnetar sight-lines are frequently in the plane of the Galaxy where even NIR extinctions can be large. However, this cannot reconcile the lower PL indices of the new counterparts with previously established ones: correcting for NIR extinction would only increase the NIR flux and decrease the indices, pushing them further from the existing population. Assuming that {\it intrinsic} indices of $0.5-1$ are typical, this suggests that the new candidates reported here are either chance alignments or are fundamentally different in nature to the other counterparts. This could include, for instance, bound companion stars, a possibility which we will explore in a subsequent paper (Chrimes et al., in prep). 

A possible exception is AX\,J1818, for which the two candidates in the error circle both have $\Gamma$ values similar to known counterparts. The likely explanation is that one of the sources is the genuine counterpart, and the other is a brighter, bluer object at a larger distance with more extinction, such that it appears to have a similar magnitude and colour. There is a similar situation for 1RXS\,J1708, which has three similar (in terms of $\Gamma$) candidates in the error circle. To determine whether this is a likely scenario by chance, we consider both the magnitude and colour of the AX\,J1818 candidates. We first note that the P$_{\rm chance}$ values (based on F160W magnitudes only) are high for these sources, at 0.32 and 0.24. Secondly, in terms of F125W-F160W colours, the AX\,J1818 values of $\sim$1.45 are typical for the field: 17 per cent of sources in the \hst\ images have redder colours, and 83 per cent are bluer. Therefore, it is not improbable to find chance alignments within the error circle which happen to have a similar appearance to genuine counterparts. 

Although not part of this \hst\ sample, we also calculate the NIR -- X-ray spectral index for the possible Galactic FRB source SGR\,1935. Using a quiescent X-ray flux level of $\sim 5 \times 10^{-11}$\,erg\,s$^{-1}$\,cm$^{-2}$, and \hst\ measurement of m$_{\rm F140W} \sim 25$ \citep{2021arXiv211207023L}, we find that this object has $\Gamma \sim 1.2$. Although this is the highest value in the population, the steep, positive slope is similar to other established counterparts. 

At the extreme low end of the PL index distribution are SGR\,0755, possibly an unrelated HMXB as discussed, and 3XMM\,J1852, for which we simply selected the reddest source in the large error circle. If NIR - X-ray spectral indices of $0.5-1$ are typical of genuine magnetar counterparts, this suggests that simply searching for anomalously red objects may be not a reliable method of identification. 

Going forwards, spectral energy distributions constructed from photometry will be needed, if not spectra, to reliably separate magnetar counterparts from other sources in the field. This can also be seen in Figure \ref{fig:obsvcmd}, where the counterparts do not always stand out from other sources. Counterpart spectral energy distributions should be constructed using data taken as close to simultaneously as possible, given that these sources can be highly variable, as demonstrated by Figure \ref{fig:variability}.

\section{Conclusions}\label{sec:conc}
In this paper, we have measured the magnitudes and colours of Galactic magnetar counterpart candidates in deep \hst\ imaging, adding a later epoch for several sources, and confirming their variability on 5-10 year time-scales. We identify six new NIR counterpart candidates for SGR\,J1622-4950, Swift\,J1822.3-1606, CXOU\,J171405.7-381031, Swift\,J1833-0832, Swift\,J1834.9-0846 and AX\,J1818.8-1559. This represents a substantial $\sim$50 per cent increase in the NIR counterpart sample size. Placing these new candidates in the context of the wider population, we find that they have typical apparent magnitudes but lower NIR - X-ray indices, with the exception of the AX\,J1818 candidates. This implies either that the new candidates are chance alignments, or that the emission mechanisms are distinct from previously established counterparts. To make further progress in identifying and understanding these counterparts, more data points are needed across the optical/NIR, since two-band colours are not necessarily enough to distinguish them from other sources. It is important for such observations to be taken as close together in time as possible, as magnetar counterparts can be highly variable over time-scales of months and years, and the inferred properties could vary substantially if different epochs are combined.

\section*{Acknowledgements}
AAC is supported by the Radboud Excellence Initiative. AJL has received funding from the European Research Council (ERC) under the European Union’s Seventh Framework Programme (FP7-2007-2013) (Grant agreement No. 725246). PJG acknowledges support from the NRF SARChI program under grant number 111692. JDL acknowledges support from a UK Research and Innovation Fellowship (MR/T020784/1). CK and ASF acknowledge support for this research, provided by NASA through a grant from the Space Telescope Science Institute, which is operated by the Association of Universities for Research in Astronomy, Inc. 

Support for this research was provided by NASA through a grant from the Space Telescope Science Institute, which is operated by the Association of Universities for Research in Astronomy, Inc. Observations analysed in this work were taken by the NASA/ESA Hubble Space Telescope under programs 14805, 15348 and 16019.  

This work has made use of {\sc ipython} \citep{2007CSE.....9c..21P}, {\sc numpy} \citep{2020arXiv200610256H}, {\sc scipy} \citep{2020NatMe..17..261V}; {\sc matplotlib} \citep{2007CSE.....9...90H},{\sc astropy},\footnote{https://www.astropy.org} a community-developed core Python package for Astronomy \citep{astropy:2013, astropy:2018} and {\sc photutils}, an Astropy package for detection and photometry of astronomical sources \citep{larry_bradley_2021_4624996}. We have also made use Seaborn packages \citep{Waskom2021}.

This research has made use of the SVO Filter Profile Service (\url{http://svo2.cab.inta-csic.es/theory/fps/}) supported from the Spanish MINECO through grant AYA2017-84089 \citep{2012ivoa.rept.1015R,2020sea..confE.182R}. 

This research has made use of software provided by the Chandra X-ray Center (CXC) in the application of the CIAO package. 

This publication makes use of data products from the Two Micron All Sky Survey, which is a joint project of the University of Massachusetts and the Infrared Processing and Analysis Center/California Institute of Technology, funded by the National Aeronautics and Space Administration and the National Science Foundation.

This work has made use of data from the European Space Agency (ESA) mission {\it Gaia} (\url{https://www.cosmos.esa.int/gaia}), processed by the {\it Gaia} Data Processing and Analysis Consortium (DPAC,
\url{https://www.cosmos.esa.int/web/gaia/dpac/consortium}). Funding for the DPAC has been provided by national institutions, in particular the institutions participating in the {\it Gaia} Multilateral Agreement.

Finally, we thank the anonymous referee for their constructive comments regarding the astrometry and clarity of this manuscript.

\section*{Data Availability}
Based on observations made with the NASA/ESA Hubble Space Telescope, obtained from the data archive at the Space Telescope Science Institute. STScI is operated by the Association of Universities for Research in Astronomy, Inc. under NASA contract NAS 5-26555. These observations are associated with programs 14805, 15348 and 16019 (Levan). The scientific results reported in this article are based on data obtained from the Chandra Data Archive.




\bibliographystyle{mnras}
\bibliography{hstcounterparts} 








\bsp	
\label{lastpage}
\end{document}